\documentclass[]{aa}
\bibpunct{(}{)}{;}{a}{}{,}

\usepackage{graphicx,txfonts,lscape,natbib}

\begin{document}
\newcommand{\APer}{$\alpha$\,Per}
\newcommand{\gbp}{{G_\mathrm{BP}}}
\newcommand{\grp}{{G_\mathrm{RP}}}

   \title{Search for the sub-stellar lithium depletion boundary in the open star cluster Coma Berenices}


   \author{E.\ L.\ Mart\'in \inst{1,2,3}
       \and
       N.\ Lodieu \inst{1,2}
       \and
       V.\ J.\ S.\ B\'ejar \inst{1,2}
        }

   \institute{
       Instituto de Astrof\'isica de Canarias, Calle V\'ia L\'actea s/n, E-38200 La Laguna, Tenerife, Spain
       \and
       Departamento de Astrof\'isica, Universidad de La Laguna, E-38206 La Laguna, Tenerife, Spain
       \and
       Consejo Superior de Investigaciones Científicas, E-28006 Madrid, Spain
       }


 
  \abstract
   {}
   {We mainly aim to search for the lithium depletion boundary (LDB) among the sub-stellar population of the open star cluster Coma Berenices.
    }
   { Since the number of brown dwarf candidates in Coma Ber available in the literature is scarce, we carried out a search for additional candidates photometrically using colour-magnitude diagrams combining
   optical and infrared photometry from the latest public releases of the following large-scale surveys: the United Kingdom InfraRed Telescope Infrared Deep Sky Survey (UKIRT/UKIDSS), the Panoramic Survey Telescope and Rapid Response System (Pan-STARRS), the Sloan Digital Sky Survey (SDSS), and AllWISE. We checked astrometric consistency with cluster membership using $Gaia$ DR2. A search for Li in three new and five previously known brown dwarf candidate cluster members was performed via spectroscopic observations using the OSIRIS instrument at the 10.4 m Gran Telescopio de Canarias (GTC). 
   }
   { A couple dozen new photometric candidate brown dwarfs located inside the tidal radius of Coma Ber are reported, but none of these are significantly fainter and cooler than previously known members. No LiI resonance doublet at 6707.8\,\AA{} 
 was detected in any of eight Coma Ber targets in the magnitude range J=15--19 and G=20--23 observed with the GTC. Spectral types and radial velocities were derived from the GTC spectra. These values confirm the cluster membership of four L2--L2.5 dwarfs, two of which are new in the literature\@. 
   }
   {The large Li depletion factors found among the four bona fide sub-stellar members in Coma Ber implies that the LDB must be located at spectral type later than L2.5 in this cluster. Using the latest evolutionary models for brown dwarfs, a lower limit of 550 Myr on the cluster age is set. This constraint has been combined with other dating methods to obtain an updated age estimate of 780$\pm$230 Myr for the Coma Ber open cluster. Identification of significantly cooler sub-stellar cluster members in Coma Ber awaits the advent of the Euclid wide survey, which should reach a depth of about J=23; this superb sensitivity will make it possible to determine the precise location of the sub-stellar LDB in this cluster and to carry out a complete census of its sub-stellar population. 
}


  \authorrunning{Martin et al.}
  \titlerunning{A search for the LDB in Coma Ber}

   \maketitle
%

%
%
\section{Introduction}
\label{ComaBer:intro}

Characterization of brown dwarfs (BD) in open clusters is useful to test models of substellar evolution and to provide benchmarks to constrain ages and masses of BD candidates in the solar neighborhood. 
Particular interest has been devoted to the study of the 
most stable isotope of lithium (Li$^7$), which is destroyed in stellar interiors at temperatures above 2.5$\times$10$^6$ K, i.e. before settling on the main-sequence, but not in BDs with masses below  about 0.06 M$_{\odot}$ \citep{1991MmSAI..62..171P,magazzu91,rebolo92}, and hence detection of the Li\,{\small{I}} resonance doublet can be used to break degeneracies in the H-R diagram between stellar and substellar mass objects  \citep{magazzu93,martin94b}.
In a coeval population, such as those of open clusters, the Li depletion boundary (LDB) provides a reliable dating method that has been used quite extensively in young open clusters and associations, such as for example the Pleiades \citep{stauffer98} and IC 2602 \citep{dobbie10}. 
 
 Even though the range of applicability of the LDB method was thought to be restricted to ages between 20 and 200 Myr \citep{Burke04,Juarez14}, it has been shown recently that it can be extended to older clusters such as the Hyades, for which an LDB age of 650$\pm$70 Myr has been obtained \citep{lodieu18a,martin18a}. The upper age limit to the application of the LDB in open clusters may take place at ages older than 1 Gyr when the Li brown dwarfs become so cold that atomic Li is no longer present in their atmospheres because it is locked in molecules such as LiH. So far, the coolest brown dwarfs for which a Li detection has been reported \citep{faherty14a,lodieu15b} is the T0 secondary of the nearest brown dwarf binary \citep{luhman13a}.

 Coma Berenices (hereafter abridged to Coma Ber, alternative name Melotte 111) is the second nearest open cluster to the Sun \citep[86 pc;][]{vanLeeuwen09,tang18a}. Updated information about the cluster properties, including the identification of a leading and a trailing trail extending 50 pc away from the cluster center can be found in \citet{tang19a}. Coma Ber has a well defined core with a tidal radius of 6.9 pc, an elongated shape, tidal trails, and it is on its way to mixing with a nearby younger moving group and located 60 pc away from the cluster center \citep{furnkranz19a, tang19a}. Coma Ber has solar metallicity and negligible reddening in the line of sight \citep{taylor06b}. The age has not been well determined, a wide range of ages are found in the literature \citep[300--1000 Myr;][]{tsvetkov89a}.   

Despite ongoing disruption, mass seggregation, and the likely large loss of former members, 
Coma Ber still has about a hundred of low-mass stars \citep{kraus07d}, and it appears to have retained even some BDs still bound to the cluster. Candidate BD members in Coma Ber have been identified by \citet{casewell06}. Follow-up spectra have been presented in \citet{casewell14a} and their masses have been estimated to lie between 0.07 and 0.05 M$_{\odot}$, which straddle the range of masses where the LDB is expected to be located.
Recently, \citet{tang18a} used the parallaxes from the second data release of $Gaia$
\citep{gaia18} to produce a full stellar census of Coma Ber and 
confirm one of the candidates in \citet{casewell14a}
as a bona-fide member. They identified two new substellar member candidates for which they assign
tentative spectral types of L2 and L4 from low-resolution near-infrared spectra. These authors estimate a cluster age of 800 Myr from isochrone fitting of massive cluster members, an age significantly older than has been previously adopted in the literature \citep[400--500 Myr;][]{kraus07d}. 

In this paper, we present a photometric search for new faint objects in the core of Coma Ber, aimed at increasing the substellar population for LDB determination, and a spectroscopic search for Li among the coolest confirmed cluster members.
In Section \ref{ComaBer:phot_search} we describe a photometric search of very low-mass (VLM) member candidates of Coma Ber cross-matching optical and near-infrared public surveys.
In Section \ref{ComaBer:GTCspectro} we present low-resolution optical spectroscopy
of eight very low-mass and potential BD member candidates collected with the
optical spectrograph on the 10.4-m Gran Telescopio de Canarias (GTC). Four of them are confirmed as bona-fide substellar cluster members. 
In Section \ref{ComaBer:Age_Li} we infer constraints on the age of Coma Ber from the absence of lithium in the four BD members confirmed by us and we provide a
revised cluster age combining our results with other dating methods.
 
\section{Search for very low-mass photometric candidate Coma Ber members}
\label{ComaBer:phot_search}
\begin{figure}
 \centering
  \includegraphics[width=\linewidth, angle=0]{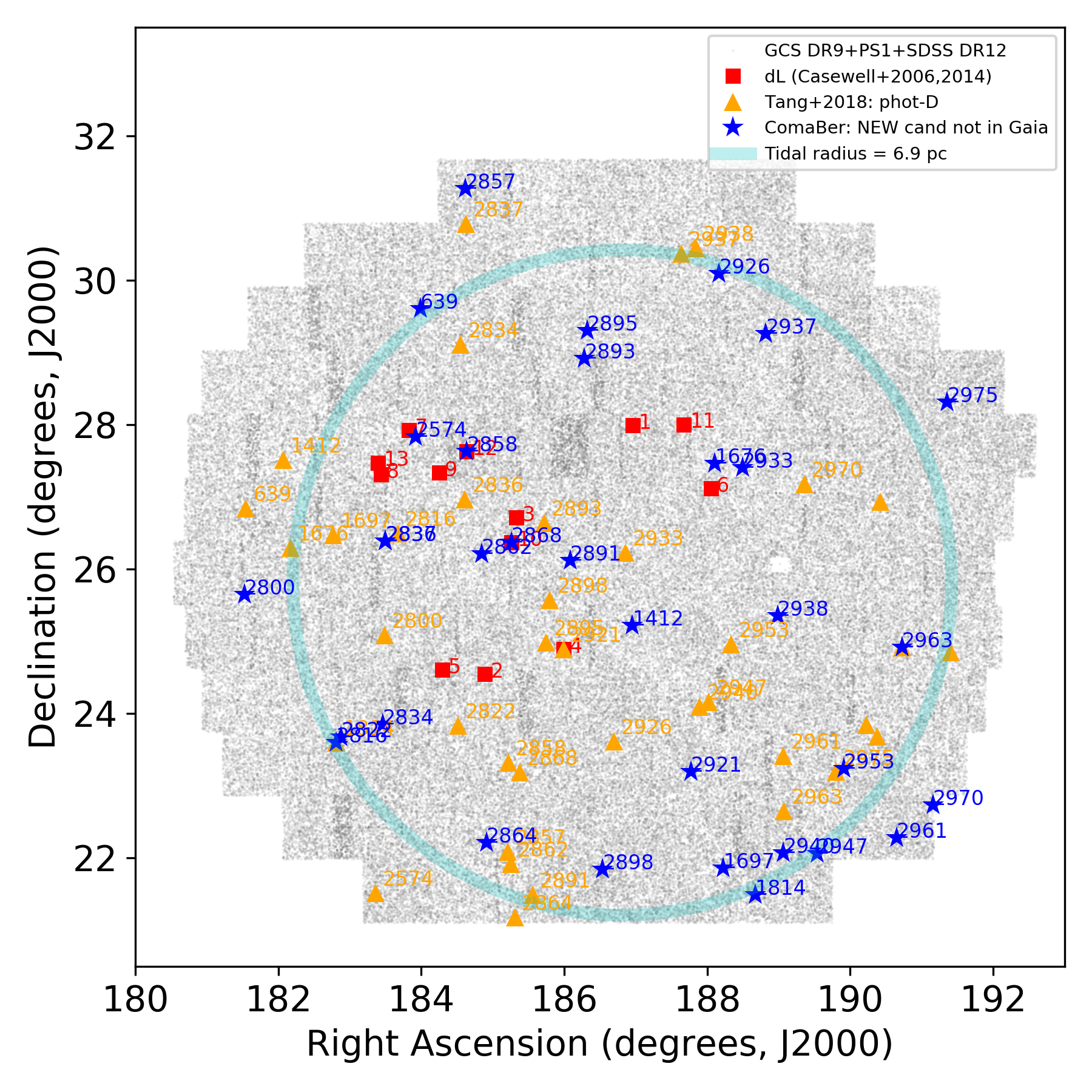}
  \caption{Coverage of the full cross-matched catalogue with UKIDSS GCS, Pan-STARRS DR1, and SDSS DR12 photometric surveys (small grey dots). Overplotted with red squares, orange triangles, and blue asterisks are member candidates of the Coma Ber cluster from \citet{casewell06} and \cite{casewell14a}, \citet{tang18a}, and this study, respectively. We included the tidal radius of Coma Ber as a large cyan circle \citep[6.9 pc;][]{tang18a}.
  }
  \label{fig_ComaBer:plot_ra_dec}
\end{figure}

 In order to increase the sample of BD targets for LDB determination in Coma Ber, 
our first step was to look for very red and faint photometric member candidates using the UKIRT Infrared Deep Sky Survey \citep[UKIDSS;][]{lawrence07} Galactic Clusters Survey (GCS) data release 9 (DR9). The search was limited to point sources fainter than 10 mag. to avoid saturation. Detections in both $J$ and $K$ bands 
with {\tt{Class}} parameters between $-$2 (probable point sources) and $-$1 (point sources) were required because L dwarfs have red $J$ and $K$ colors.
We launched the query in the UKIRT Wide Field Camera \citep[WFCAM;][]{casali07} Science Archive 
\citep{hambly08}, which returned 234905 sources over an area of 98.8 square degrees.
We show the GCS coverage in Coma Ber in Fig.\ \ref{fig_ComaBer:plot_ra_dec} where we included
the tidal radius of the cluster as a large cyan circle \citep[6.9 pc;][]{tang18a}.
The GCS coverage does not reach out to the tidal tails of Coma Ber that extend up to 50 pc
 \citep{tang18a,tang19a,furnkranz19a} but our survey is complete out to a radius of
5.6 deg centered at (ra,dec) = (186.61,+26.31) degrees.

We cross-matched this input catalogue with the first data release of the Panoramic Survey 
Telescope and Rapid Response System \citep[Pan-STARRS;][]{panstarrs13}, the Sloan Digital Sky Survey
\citep[SDSS][]{alam15a} DR12, and AllWISE \citep{wright10,cutri14} with a matching radius of 3 arcsec.
We have included the calculation of the proper motions from the baseline between Pan-STARRS and SDSS data,
whose mean value lies around 7.2 year.
The similar numbers of sources in the UKIDSS GCS DR9 catalogue and the final catalogue with 
Pan-STARRS and SDSS show that both optical surveys perfectly complement the infrared photometry 
of the UKIDSS GCS\@. We show the GCS coverage in Coma Ber in Fig.\ \ref{fig_ComaBer:plot_ra_dec}.

\begin{figure*}
 \centering
  \includegraphics[width=0.42\linewidth, angle=0]{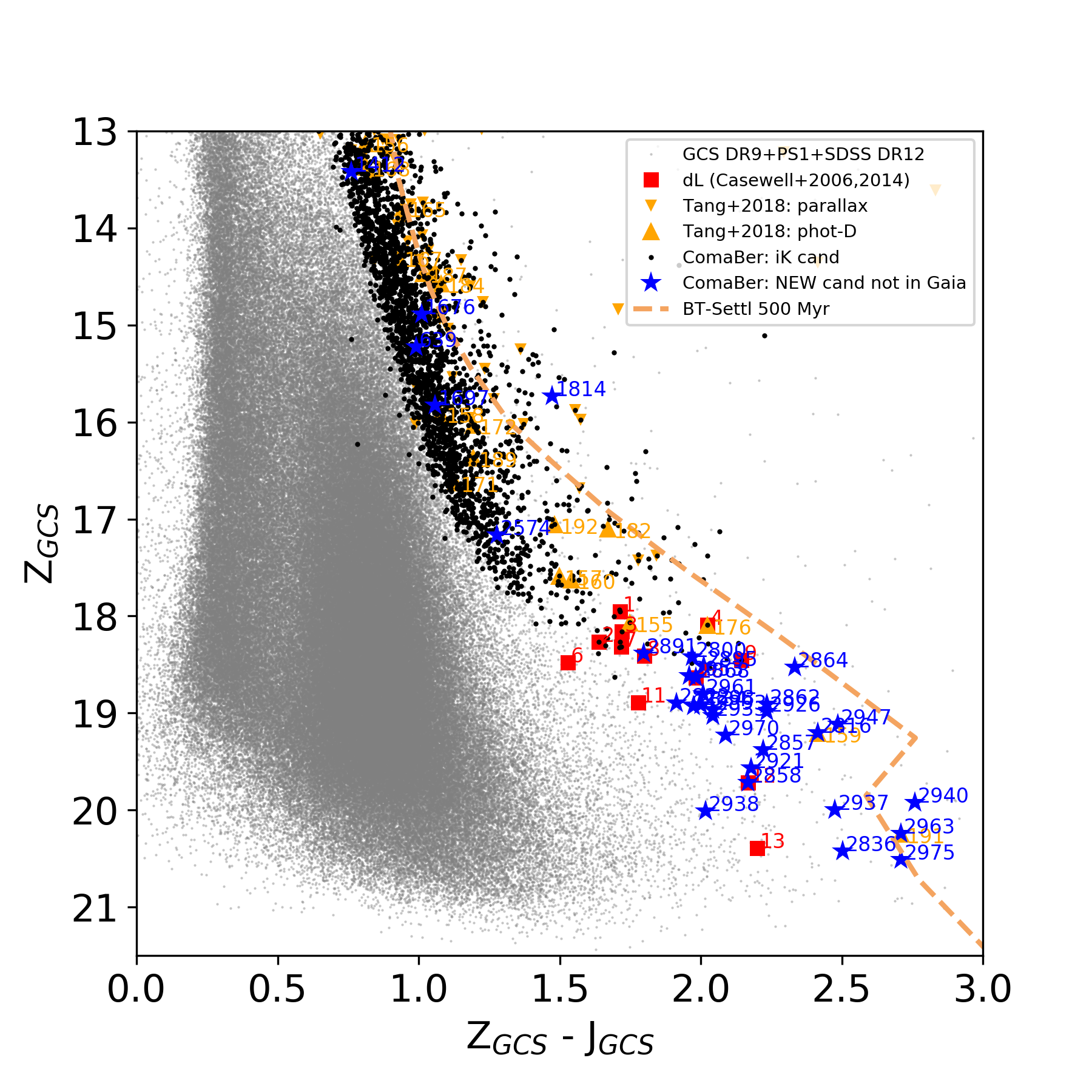}
  \includegraphics[width=0.42\linewidth, angle=0]{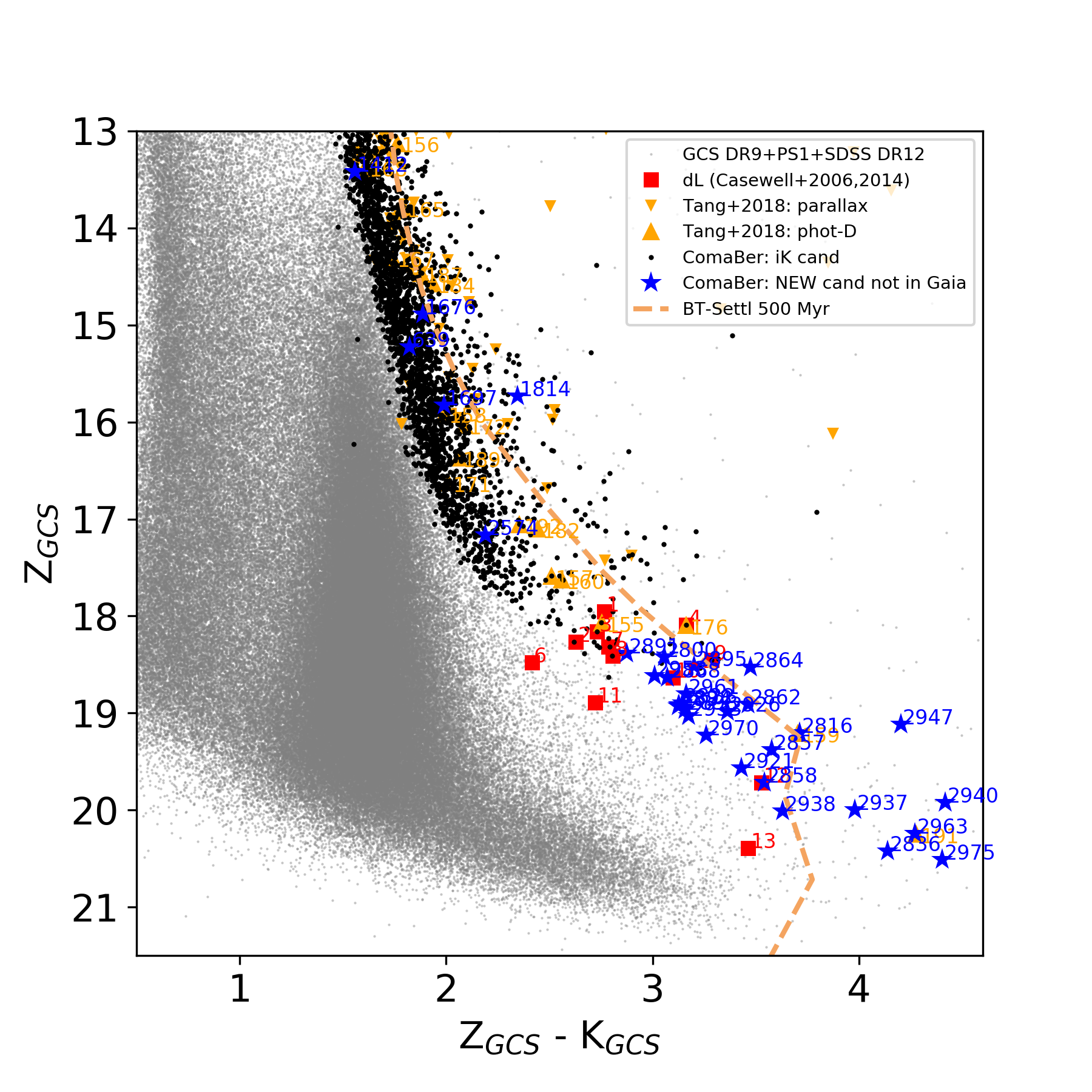}
  \includegraphics[width=0.42\linewidth, angle=0]{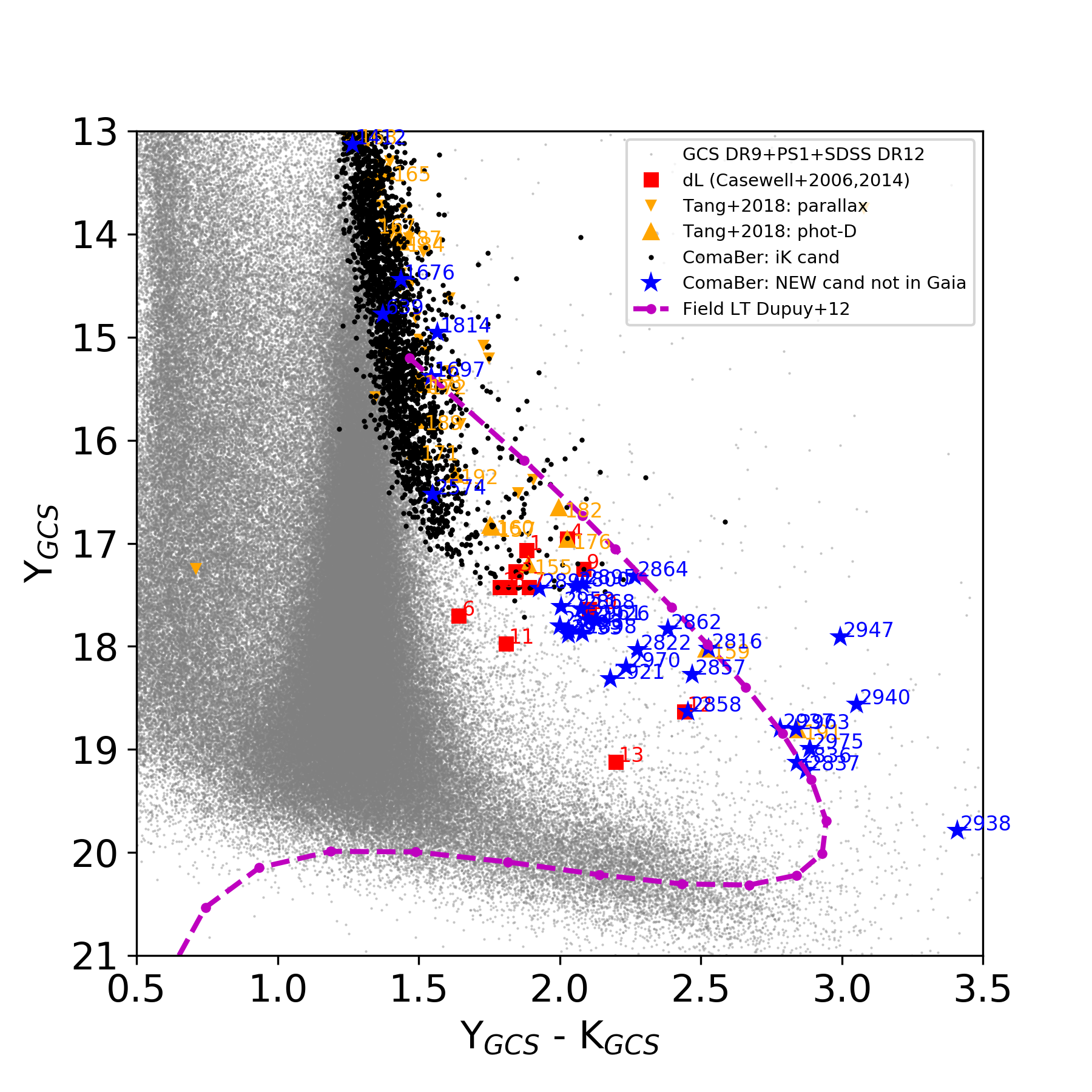}
  \includegraphics[width=0.42\linewidth, angle=0]{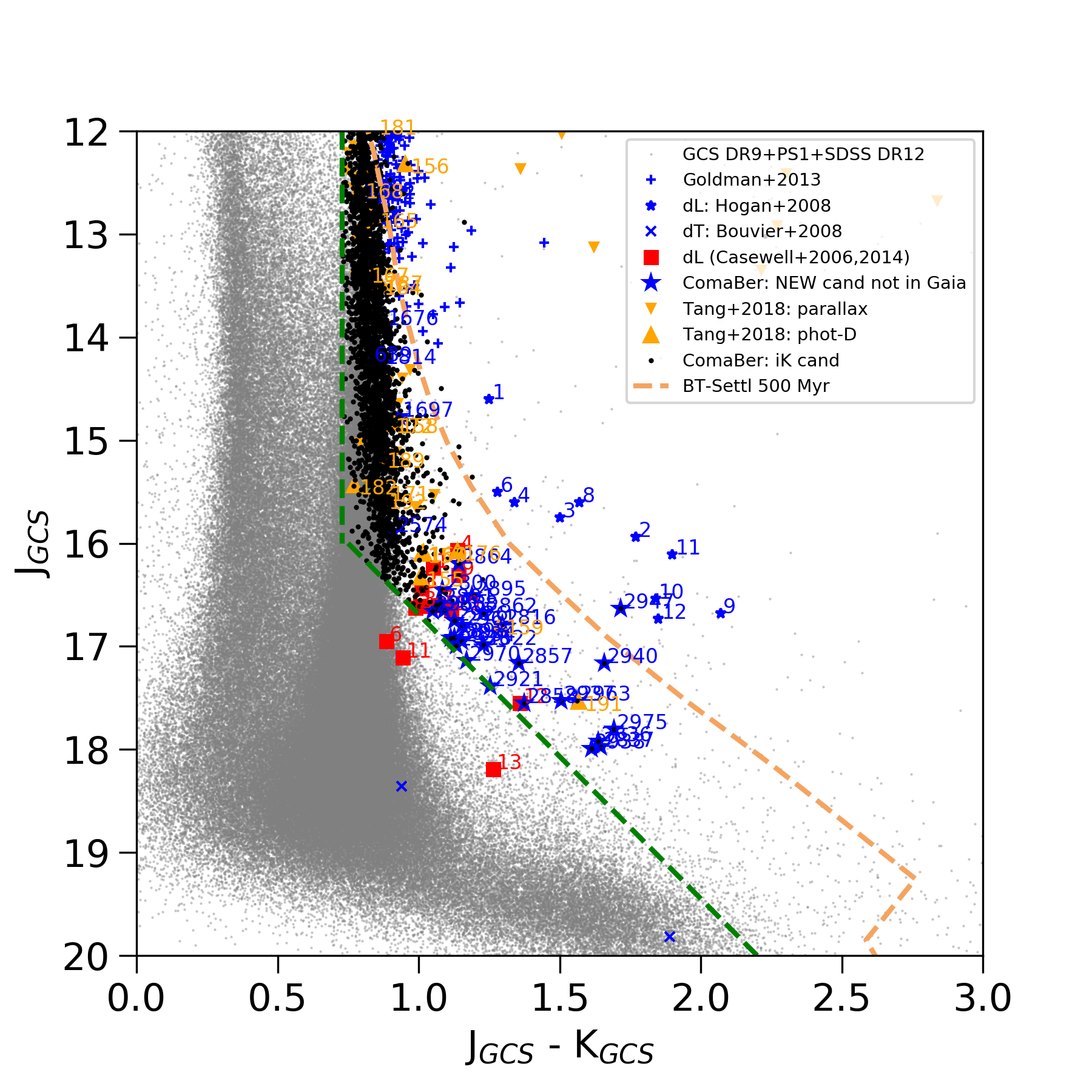}
  \includegraphics[width=0.42\linewidth, angle=0]{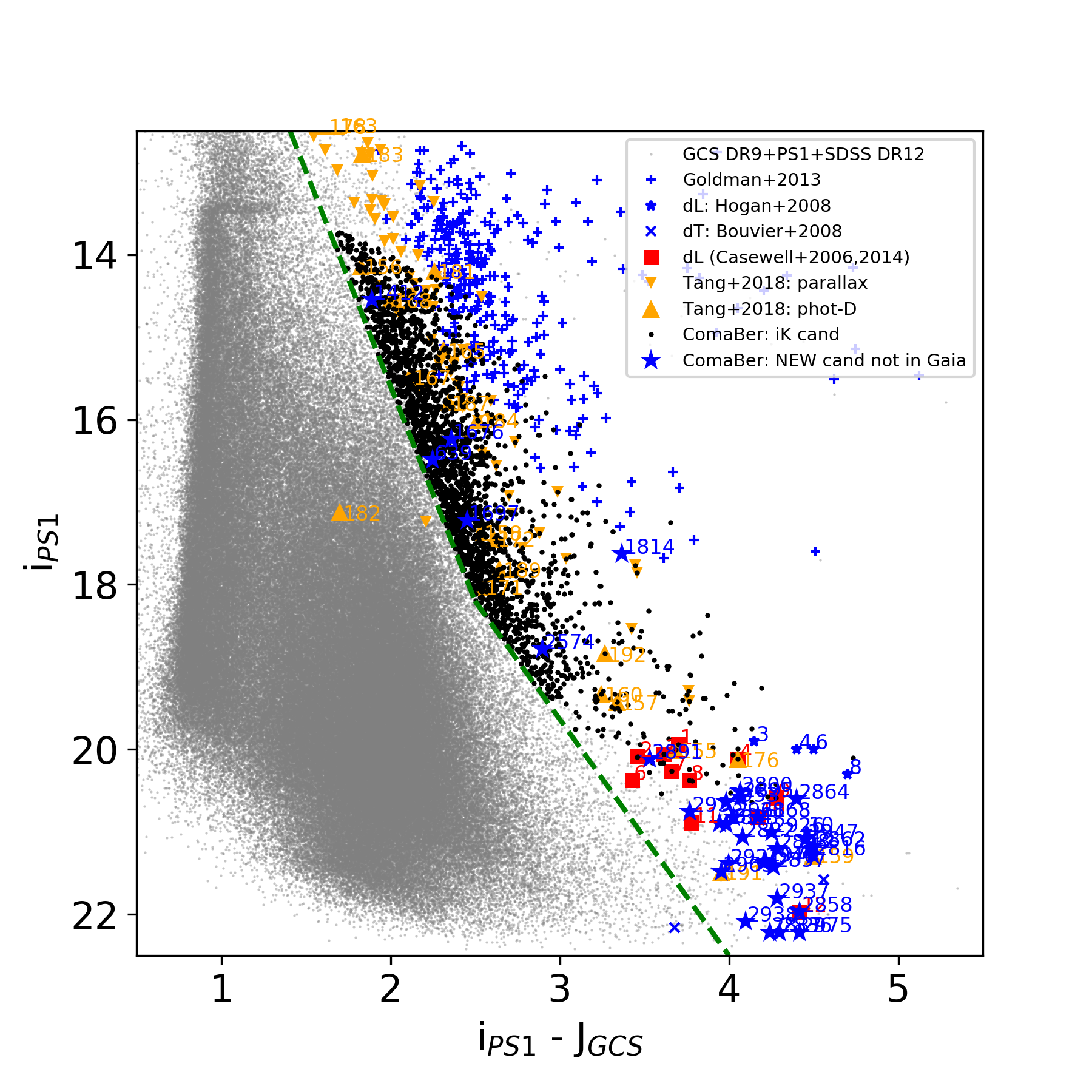}
  \includegraphics[width=0.42\linewidth, angle=0]{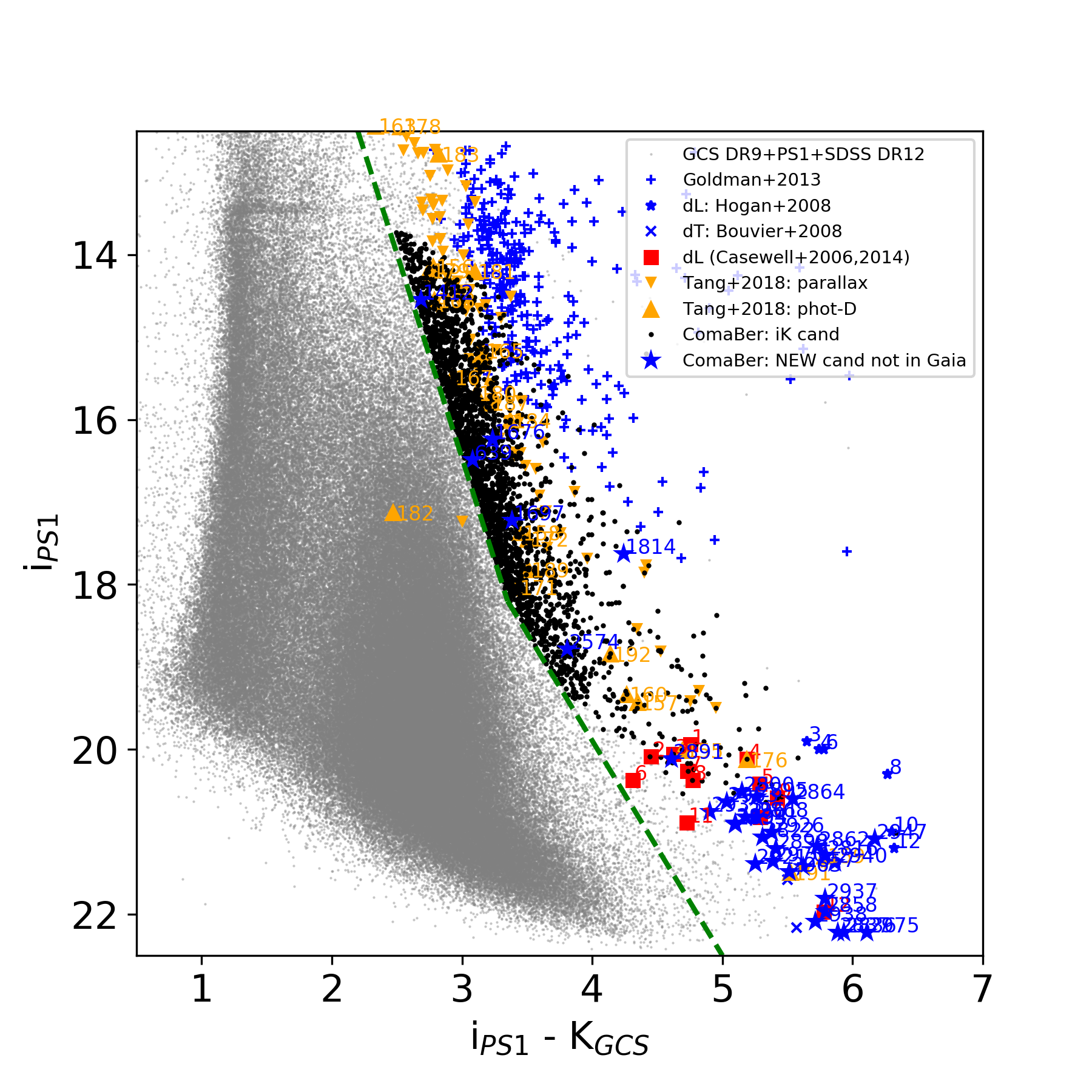}
  \caption{Colour-magnitude diagrams with optical and infrared photometry from the GCS and Pan-STARRS surveys showing all sources common to UKIDSS GCS, Pan-STARRS DR1, and SDSS DR12 (small grey dots) along with $iK$ candidates selected photometrically along the line of sight of Coma Ber. Overplotted are candidates from \citet{casewell06, casewell14a}, \citet{tang18a}, and our study as red squares, orange triangles, and blue asterisks, respectively. We added the 500 Myr BT-Settl isochrones \citep{baraffe15} as dashed orange lines and members in the Hyades \citep{hogan08,bouvier08a,goldman13}, and the sequence of field ultracool dwarfs with known distances (magenta dashed curve) for comparison in the diagrams where suitable photometric data are available. The selection cuts to identify photometric candidate members in Coma Ber are shown as green dashed lines.
}
  \label{fig_ComaBer:plot_CMD_NIR}
\end{figure*}

To define conservative photometric selection cuts, we collected known members from the surveys
of \citet{casewell06}, \citet{casewell14a}, and \citet{tang18a} and plotted in various colour-magnitude
diagrams depicted in Fig.\ \ref{fig_ComaBer:plot_CMD_NIR}. The most recent survey of \citet{tang18a}
distinguishes candidates with and without parallaxes from the $Gaia$ second data release (DR2) 
\citep{gaia18}. We excluded three candidates of \citet{casewell06} 
that lie systematically below the sequence (candidate numbers \#6, \#11, and \#13).
The cluster sequence is mainly guided by the two faintest candidates of \citet{tang18a}
with infrared spectra: T159 and T191\@.
The sequence of Hyades L dwarfs confirmed spectroscopically 
 \citep{goldman13,hogan08,bouvier08a,martin18a,lodieu18b} shifted to the distance of
ComaBer also supports the fact that the sequence of Coma Ber is marked by T159 and T191
\citep{tang18a}, which we used as reference to define our selection lines. The object \#12 
(=\,cbd67) of \citet{casewell06} is confirmed as a L2.0 member and included in our selection.
We note that the 500 Myr-old isochrone  \citep{allard12,baraffe15} and the field sequence
of old M, L, and T dwarfs \citep*{dupuy12} tend to lie above the putative sequence of
very-low mass members in Coma Ber.
 
We defined photometric selection criteria in three colour-magnitudes diagrams with a set of straight 
lines going from the brightest to the faintest limits (green dashed lines in Fig.\ \ref{fig_ComaBer:plot_CMD_NIR}:
\begin{itemize}
\item [$\bullet$] ($J-K$,$J$)\,$\geq$\,0.73 from $J$\,=\,11.2 to 16 mag
\item [$\bullet$] ($J-K$,$J$)\,=\,(0.75,16.0) to (2.20,20.0)
\item [$\bullet$] ($i_{PS1}-J$,$i_{PS1}$)\,=\,(1.35,12.2) to (2.50,18.2)
\item [$\bullet$] ($i_{PS1}-J$,$i_{PS1}$)\,=\,(2.50,18.2) to (4.00,22.5)
\item [$\bullet$] ($i_{PS1}-K$,$i_{PS1}$)\,=\,(2.20,12.2) to (3.35,18.2)
\item [$\bullet$] ($i_{PS1}-K$,$i_{PS1}$)\,=\,(3.35,18.2) to (5.00,22.5)
\end{itemize}

We started off with the ($J-K$,$J$) colour-magnitude because it is the most sensitive to red and faint cluster members. Then, we removed those not satisfying the criteria in the ($i-J$,$i$)
and ($i-K$,$i$) diagrams, yielding a list of 2583 candidates labelled as ``ComaBer: iK cand'' in
Fig.\ \ref{fig_ComaBer:plot_CMD_NIR}. We cross-matched this list with the $Gaia$ DR2 catalogue
to remove the 2550 sources with $Gaia$ information because the astrometric selection was already
done by \citet{tang18a}. We assume that this selection is complete down to the limit of $Gaia$,
corresponding approximately to $J$\,=\,16 mag. Hence, we are left with photometric candidates  
labelled as ``ComaBer: NEW cand not in Gaia'' and displayed as blue asterisks in 
Fig.\ \ref{fig_ComaBer:plot_CMD_NIR}. We should emphasise that our selection includes cbd10
and cbd67 as photometric candidates \citep{casewell14a} as well as T159 and T191 \citep{tang18a}.

\begin{figure}
 \centering
  \includegraphics[width=\linewidth, angle=0]{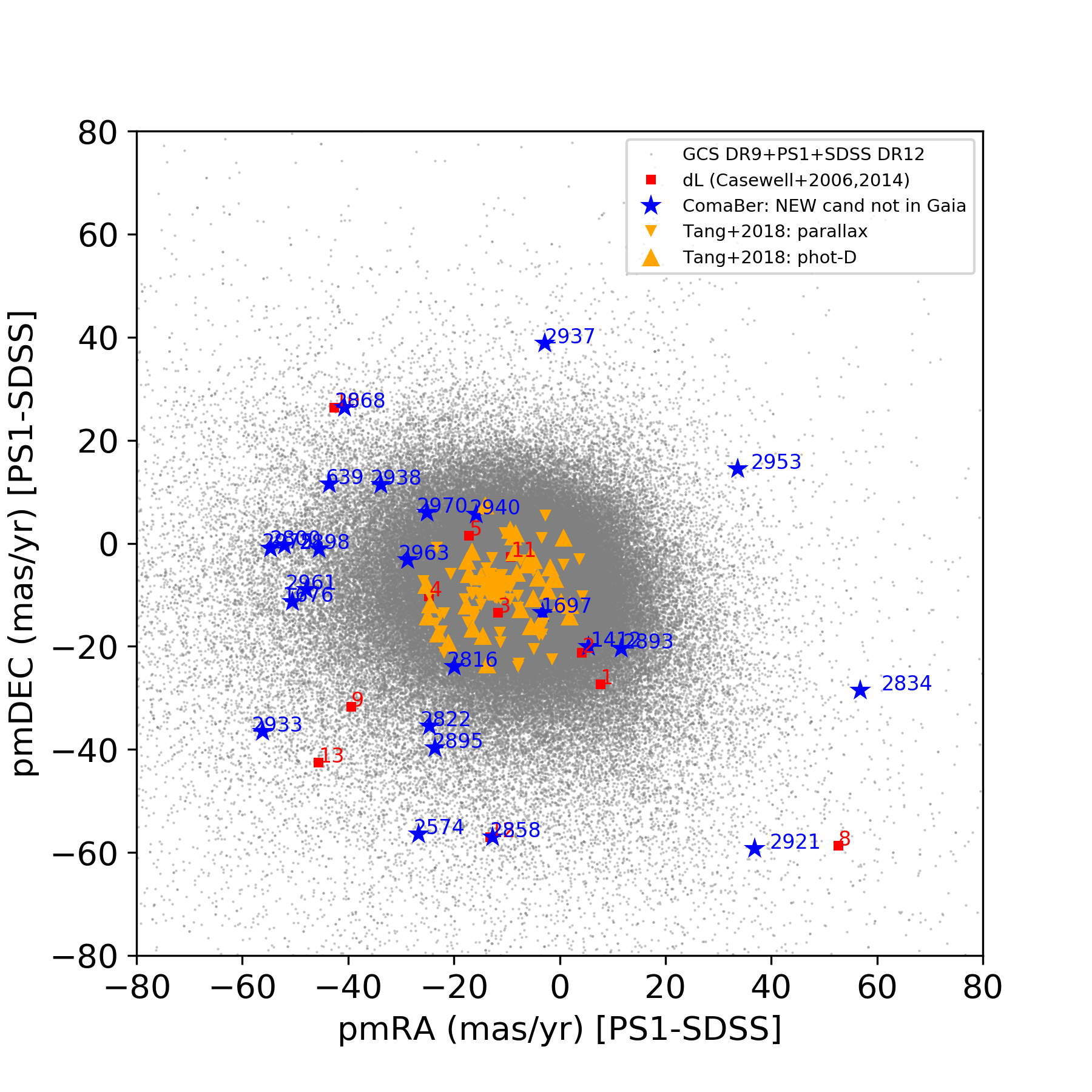}
  \caption{Vector point diagram depicting the proper motions from the Pan-STARRS DR1, vs SDSS DR12 cross-match for all sources towards Coma Ber (grey dots). Overplotted are the photometric candidates selected in this work along with candidates from the literature \citep{casewell14a,tang18a}, as in Fig.\ \ref{fig_ComaBer:plot_ra_dec}.
}
  \label{fig_ComaBer:plot_ComaBer_VPD}
\end{figure}

We computed the proper motions from the difference between the PS1 and SDSS positions, taking into
account the epoch difference (mean or median around 7.25 years). The uncertainties on the proper
motion is taken has 1.48 times the median absolute deviation (MAD). The MAD is 8.9 and 7.9 mas/yr,
yielding proper motion uncertainties of 13.2 and 11.7 mas/yr in RA and dec, respectively.
We analyse the position of these new photometric member candidates in the vector point diagram 
plotted in Fig.\ \ref{fig_ComaBer:plot_ComaBer_VPD}. We observe a large dispersion in the 
proper motions but find some sources with motions consistent with the bulk of parallax and
photometric member candidates in \citet{tang18a}, e.g.\ \#2963\,=\,T191,  \#2816\,=\,T159, 
\#2938\,=\,ComaBer4\@.
If we apply a 3$\sigma$ selection based on the uncertainties taken as 1.48$\times$MAD and centered 
on the mean motion of the cluster ($-$11.2,$-$9.2 mas/yr), we would keep eight sources out of the 
33 photometric candidates previously selected (\#2963, 2893, 2816, 1412, 1697, 2938, 2940, 2970).

We compile this list of member candidates in Table \ref{tab_ComaBer:table_new_cand}, dividing 
it up into three sub-samples: (1) 4 sources previously reported by other studies (top panel), 8 
photometric candidates with proper motion consistent with the mean motion of the cluster 
(middle panel), and the remaining 22 photometric candidates (bottom panel).

%
 
\begin{table*}
\tiny
\footnotesize
 \centering
 \caption[]{List of Coma Ber photometric member candidates identified in the area common
to the three large-scale public surveys considered in this work. We list the ID number, the
coordinates in sexagesimal format from UKIDSS GCS, and the infrared and optical photometry
from UKIDSS GCS DR9, Pan-STARRS DR1, and SDSS DR12\@.
{\it{Top panel:}} Four photometric candidates in common with earlier studies \citep{casewell14a,tang18a}.
{\it{Middle panel:}} Eight photometric candidates with proper motion consistent with the mean motion
of the cluster.
{\it{Bottom panel:}} The remaining 22 photometric candidates.
Notes: Known sources are: \#2963\,=\,T191, \#2816\,=\,T159, \#2858\,=\,cbd67, and \#2868\,=\,cbd10 .
Notes: New objects with GTC spectroscopy are: ComaBer\,4\,=\,\#2938, ComaBer\,5\,=\,\#2975, and ComaBer\,7\,=\,\#2857\@.
}
 \begin{tabular}{@{\hspace{0mm}}l @{\hspace{1mm}}c @{\hspace{1mm}}c @{\hspace{1mm}}c @{\hspace{1mm}}c 
 @{\hspace{1mm}}c @{\hspace{1mm}}c @{\hspace{1mm}}c @{\hspace{1mm}}c @{\hspace{1mm}}c @{\hspace{1mm}}c @{\hspace{1mm}}c @{\hspace{1mm}}c @{\hspace{1mm}}c @{\hspace{1mm}}c @{\hspace{1mm}}c@{\hspace{0mm}}}
 \hline
 \hline
ID &    R.A.     &   Dec      &   $Z$ & $Y$ & $J$ & $H$ & $K$ & 
PS$i$ & PS$z$ & PS$y$ & SDSS$u$ & SDSS$g$ & SDSS$r$ & SDSS$i$ & SDSS$z$  \cr
 \hline
   & hh:mm:ss.s & ${^\circ}$:$'$:$''$ & mag & mag & mag & mag 
   & mag & mag & mag & mag & mag & mag & mag & mag & mag \cr
 \hline
2816 & 12:11:14.9 & 23:35:39.9 & 19.202 & 18.019 & 16.787 & 16.109 & 15.471 & 
21.284 & 19.829 & 18.831 & 24.166 & 25.065 & 23.147 & 21.197 & 19.602 \cr
2858 & 12:18:32.7 & 27:37:30.8 & 19.719 & 18.632 & 17.551 & 16.779 & 16.183 &  
21.970 & 20.508 & 19.476 & 26.364 & 25.419 & 23.700 & 22.328 & 20.154 \cr
2868 & 12:21:02.4 & 26:22:04.3 & 18.633 & 17.638 & 16.651 & 16.082 & 15.553 & 
20.822 & 19.368 & 18.448 & 23.151 & 24.653 & 25.598 & 20.721 & 19.046 \cr
2963 & 12:42:53.7 & 24:55:07.1 & 20.240 & 18.807 & 17.530 & 16.688 & 15.924 &  
21.485 & 20.674 & 19.719 & 23.854 & 24.751 & 24.571 & 22.577 & 20.381 \cr
\hline
1412 & 12:27:48.3 & 25:13:11.0 & 13.420 & 13.127 & 12.658 & 12.131 & 11.798 & 
14.548 & 14.112 & 13.876 & 19.335 & 16.938 & 15.531 & 14.911 & 13.981 \cr
1697 & 12:32:54.2 & 21:51:17.1 & 15.830 & 15.384 & 14.771 & 14.183 & 13.855 & 
17.224 & 16.480 & 16.108 & 22.789 & 20.356 & 18.807 & 17.208 & 16.306 \cr
2816 & 12:11:14.9 & 23:35:39.9 & 19.202 & 18.019 & 16.787 & 16.109 & 15.471 & 
21.284 & 19.829 & 18.831 & 24.166 & 25.065 & 23.147 & 21.197 & 19.602 \cr
2893 & 12:25:06.9 & 28:54:48.0 & 18.966 & 17.839 & 16.923 & 16.361 & 15.860 & 
20.907 & 19.533 & 18.629 & 25.068 & 25.042 & 24.801 & 20.936 & 19.313 \cr
2938 & 12:35:56.6 & 25:21:11.2 & 20.010 & 19.789 & 17.993 & 17.127 & 16.400 &  
22.091 & 20.956 & 20.073 & 26.337 & 24.222 & 24.289 & 21.949 & 20.464 \cr
2940 & 12:36:15.9 & 22:03:45.7 & 19.924 & 18.560 & 17.165 & 16.292 & 15.564 & 
21.370 & 20.483 & 19.347 & 24.868 & 25.350 & 25.395 & 22.100 & 20.033 \cr
2963 & 12:42:53.7 & 24:55:07.1 & 20.240 & 18.807 & 17.530 & 16.688 & 15.924 &  
21.485 & 20.674 & 19.719 & 23.854 & 24.751 & 24.571 & 22.577 & 20.381 \cr
2970 & 12:44:37.8 & 22:43:41.9 & 19.228 & 18.205 & 17.139 & 16.494 & 15.971 & 
21.357 & 20.005 & 18.993 & 24.756 & 24.382 & 23.646 & 21.333 & 19.463 \cr
\hline
639  & 12:15:57.2 & 29:36:20.8 & 15.228 & 14.778 & 14.237 & 13.714 & 13.391 & 
16.486 & 15.818 & 15.505 & 22.654 & 19.573 & 18.007 & 16.500 & 15.703 \cr
1676 & 12:32:25.4 & 27:27:29.3 & 14.891 & 14.441 & 13.881 & 13.335 & 13.011 & 
16.240 & 15.530 & 15.175 & 21.957 & 19.357 & 17.831 & 16.226 & 15.352 \cr
1814 & 12:34:40.5 & 21:29:25.8 & 15.732 & 14.953 & 14.258 & 13.797 & 13.397 & 
17.627 & 16.390 & 15.770 & 24.560 & 21.886 & 20.452 & 17.634 & 16.197 \cr
2574 & 12:15:41.2 & 27:49:42.9 & 17.167 & 16.527 & 15.888 & 15.344 & 14.955 & 
18.785 & 17.825 & 17.316 & 24.673 & 22.411 & 20.890 & 18.795 & 17.631 \cr
2800 & 12:06:06.0 & 25:38:35.5 & 18.420 & 17.422 & 16.449 & 15.857 & 15.359 & 
20.516 & 19.109 & 18.236 & 24.163 & 24.698 & 23.804 & 20.648 & 18.871 \cr
2822 & 12:11:30.6 & 23:40:16.5 & 18.899 & 18.032 & 16.986 & 16.331 & 15.792 & 
21.066 & 19.667 & 18.773 & 23.700 & 24.499 & 23.959 & 21.119 & 19.322 \cr
2834 & 12:13:50.1 & 23:50:42.7 & 18.926 & 17.801 & 16.953 & 16.314 & 15.803 & 
20.897 & 19.488 & 18.564 & 24.951 & 25.181 & 24.520 & 20.624 & 19.147 \cr
2836 & 12:13:59.8 & 26:23:14.8 & 20.422 & 19.126 & 17.919 & 17.141 & 16.296 &  
22.218 & 20.954 & 19.826 & 25.643 & 24.824 & 24.798 & 21.867 & 20.765 \cr
2857 & 12:18:27.2 & 31:15:53.3 & 19.382 & 18.276 & 17.162 & 16.357 & 15.767 &  
21.427 & 20.085 & 19.241 & 23.334 & 23.950 & 23.602 & 21.355 & 19.808 \cr
2862 & 12:19:22.0 & 26:12:30.7 & 18.909 & 17.830 & 16.675 & 16.071 & 15.467 &  
21.175 & 19.604 & 18.700 & 25.299 & 25.986 & 23.306 & 21.251 & 19.479 \cr
2864 & 12:19:38.4 & 22:12:36.2 & 18.531 & 17.321 & 16.198 & 15.570 & 15.013 & 
20.599 & 19.142 & 18.155 & 25.500 & 25.154 & 22.837 & 20.410 & 18.561 \cr
2891 & 12:24:19.6 & 26:06:53.6 & 18.383 & 17.436 & 16.586 & 15.951 & 15.470 &  
20.119 & 18.961 & 17.927 &  ---   &  ---   &  ---   &  ---   &  ---   \cr
2895 & 12:25:16.4 & 29:18:04.2 & 18.518 & 17.396 & 16.506 & 15.847 & 15.276 & 
20.578 & 19.210 & 18.322 & 23.374 & 22.171 & 21.348 & 19.873 & 18.152 \cr
2898 & 12:26:08.2 & 21:50:10.9 & 18.917 & 17.872 & 16.917 & 16.326 & 15.816 & 
21.203 & 19.625 & 18.738 & 23.996 & 25.591 & 24.372 & 21.226 & 19.295 \cr
2921 & 12:31:04.5 & 23:11:42.6 & 19.566 & 18.315 & 17.388 & 16.648 & 16.130 &  
21.386 & 19.949 & 19.099 & 25.349 & 24.812 & 23.786 & 21.537 & 19.840 \cr
2926 & 12:32:39.4 & 30:05:15.6 & 18.987 & 17.752 & 16.752 & 16.140 & 15.659 &  
21.004 & 19.585 & 18.662 & 24.451 & 25.676 & 22.462 & 21.136 & 19.086 \cr
2933 & 12:33:58.2 & 27:24:12.6 & 19.032 & 17.887 & 16.989 & 16.335 & 15.816 &  
20.759 & 19.520 & 18.667 & 23.533 & 24.009 & 23.446 & 21.041 & 19.158 \cr
2937 & 12:35:16.6 & 29:15:33.3 & 20.001 & 18.801 & 17.525 & 16.766 & 16.143 &  
21.810 & 20.590 & 19.638 & 21.339 & 28.902 & 20.789 & 19.579 & 17.772 \cr
2947 & 12:38:09.0 & 22:03:32.2 & 19.116 & 17.909 & 16.631 & 15.684 & 14.895 &  
21.086 & 19.738 & 18.839 & 25.316 & 25.322 & 23.387 & 21.183 & 19.364 \cr
2953 & 12:39:37.9 & 23:14:08.4 & 18.617 & 17.614 & 16.657 & 16.135 & 15.660 & 
20.641 & 19.255 & 18.400 & 22.850 & 23.734 & 21.706 & 20.168 & 18.671 \cr
2961 & 12:42:35.8 & 22:16:43.9 & 18.806 & 17.746 & 16.797 & 16.134 & 15.688 & 
20.822 & 19.446 & 18.524 & 25.025 & 24.623 & 23.428 & 20.966 & 19.123 \cr
2975 & 12:45:25.2 & 28:18:16.2 & 20.512 & 18.995 & 17.801 & 16.946 & 16.145 &  
22.219 & 20.726 & 19.603 & 24.614 & 24.995 & 24.860 & 22.175 & 20.436 \cr
 \hline
\label{tab_ComaBer:table_new_cand}
\end{tabular}
\end{table*}

%

\section{Spectroscopic observations and analysis}
\label{ComaBer:GTCspectro}

\subsection{Low-resolution optical spectroscopy}
\label{ComaBer:GTCspectro_obs_DR}

We obtained optical spectra with the Optical System for Imaging and low Resolution Integrated Spectroscopy
\citep[OSIRIS;][]{cepa00} instrument on the 10.4\,m Gran Telescopio de Canarias (GTC) over two semesters
(GTC47-18A and GTC55-19A; PI Mart\'in). 
All of the data were collected under service mode, except four sources (ComaBer\,4, 5, 7, and 13) which were observed during a 2-night visitor mode campaign on 12 and 13 May 2018
(observer V.J.S. B\'ejar). 
The log of the observations is available in Table \ref{tab_ComaBer:log_obs}.
We requested dark conditions, seeing better than 1.2 arcsec, and spectroscopic conditions for both GTC programes.

We obtained low-resolution optical spectra with the R1000R grating covering the 5100--10000\,\AA{} wavelength range and a slit of 1.2 arcsec, yielding a spectral resolving power of about 560 at the central wavelength. We acquired our targets in parallactic angle with the Sloan $z$-band filter due to their faintness in the optical. We employed on-source integrations with the individual exposure times provided in Table \ref{tab_ComaBer:log_obs}, 
with 2 to 4 repeats shifted along the slit by 10 to 15 arcsec depending on the crowding of the field to allow for sky subtraction. As part of programme GTC55-19A, we also observed a nearby brown dwarf spectral standard with a clear lithium absorption, namely DENIS\,1228 \citep{martin97b,martin99a}. 
 
\begin{table*}
 \centering
 \caption[]{Log of the GTC OSIRIS spectroscopic observations.
We list the names of the targets as given in \citet{casewell06} and \citet{tang18a}, and
this paper, respectively, along with their coordinates, magnitudes, dates of observations,
instrumental set-up, weather conditions, and spectral types derived in this work.
The last row is a well-known field lithium L dwarf template \citep{martin97b, martin99a} observed as part of our programme GTC55-19A\@.
}
 \begin{tabular}{@{\hspace{0mm}}l c c c c c c c@{\hspace{0mm}}}
 \hline
 \hline
Name        &    R.A.     &     Dec       &   $J$  &   $G$  &    Date   &    Date     &   SNR    \cr
 \hline
            & hh:mm:ss.ss & ${^\circ}$:$'$:$''$ & mag & mag & yyyy-mm-dd &   sec        &   6750-6800 \AA          \cr
 \hline
cbd67       & 12:18:32.71 & $+$27:37:31.3 & 17.551 & 22.485 & 13-04-2018 & 2$\times$1800 & 4.9 \cr
cbd34       & 12:23:57.37 & $+$24:53:29.0 & 15.940 & 20.140 & 13-04-2018 & 2$\times$1500 & 12.9 \cr
cbd34       & 12:23:57.37 & $+$24:53:29.0 & 15.940 & 20.140 & 13-05-2018 & 2$\times$1500 & 7.8 \cr
cbd10       & 12:21:02.46 & $+$26:22:04.2 & 16.814 & 21.139 & 12-04-2018 & 2$\times$1500 & 2.0  \cr
ComaBer\,4  & 12:35:56.59 & $+$25:21:11.2 & 17.993 & 22.928 & 13-05-2018 & 2$\times$1800 &  4.9 \cr
ComaBer\,5  & 12:45:25.19 & $+$28:18:16.2 & 17.801 & 22.736 & 13-05-2018 & 6$\times$1800 & 7.0 \cr
ComaBer\,7  & 12:18:27.25 & $+$31:15:53.3 & 17.162 & $>$21.36 & 12-05-2018 & 2$\times$1500 & 3.6 \cr
T159        & 12:11:14.90 & $+$23:35:39.9 & 16.787 & 22.736 & 29-04-2018 & 2$\times$1800 & 3.9 \cr
T191        & 12:42:53.70 & $+$24:55:07.1 & 17.530 & 22.736 & 29-04-2018 & 4$\times$1960 & 4.1 \cr
T191        & 12:42:53.70 & $+$24:55:07.1 & 17.530 & 22.736 & 01-05-2018 & 2$\times$1960 & 3.2  \cr
T191        & 12:42:53.70 & $+$24:55:07.1 & 17.530 & 22.736 & 11-05-2018 & 2$\times$1960 & 7.6 \cr
 \hline
DENIS1228 & 12:28:15.20 & $-$15:47:34.2 & 14.378 &  ---   & 13-04-2018  & 2$\times$370  & 10.1  \cr
 \hline
\label{tab_ComaBer:log_obs}
\end{tabular}
\end{table*}

\begin{figure}
 \centering
\includegraphics[width=\linewidth, angle=0]{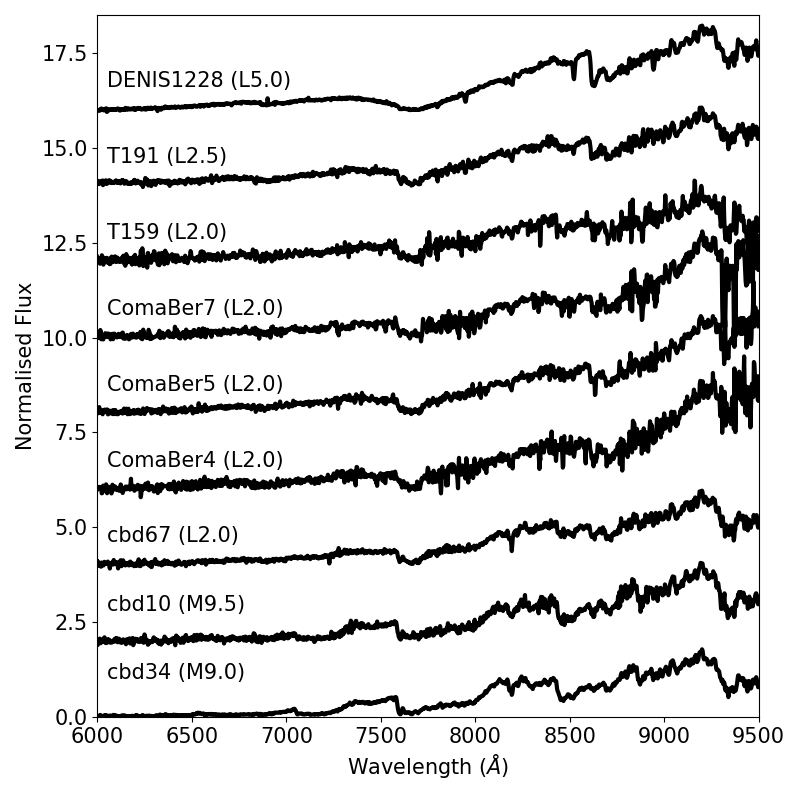}
 \caption{Final set of GTC/OSIRIS spectra discussed in this paper. Spectral types assigned to each object are labelled.}
  \label{fig_ComaBer:plot_spec_full}
\end{figure}

We reduced the data under the IRAF environment \citep{tody93} in a standard manner\footnote{IRAF is distributed by the National Optical Astronomy Observatory, which is operated by the Association of Universities for Research in Astronomy (AURA) under cooperative agreement with the National Science Foundation}.
First, we median-combined the bias and flat-fields taken during the afternoon, which we removed from the raw images. We extracted optimally each individual spectrum choosing the aperture and sky regions. We calibrated the spectra in wavelength with HgAr, Xenon, and Neon lamps with a rms better than 0.3\,\AA{} before averaging them to produce the final spectra shown in Fig.\ \ref{fig_ComaBer:plot_spec_full}.
We flux calibrated the 1D spectra with the response functions derived from two spectrophotometric standard stars: Hilt\,600 
\citep[B1;][]{pancino12} 
and Ross\,640 \citep[DZA5.5;][]{greenstein67,harrington80,wesemael93}
for programmes GTC47-18A and GTC55-19A, respectively. We did not correct for the second-order contamination so the flux calibration is solely valid up to 9500\,\AA{}. 

\begin{figure}
 \centering
 \includegraphics[width=\linewidth, angle=0]{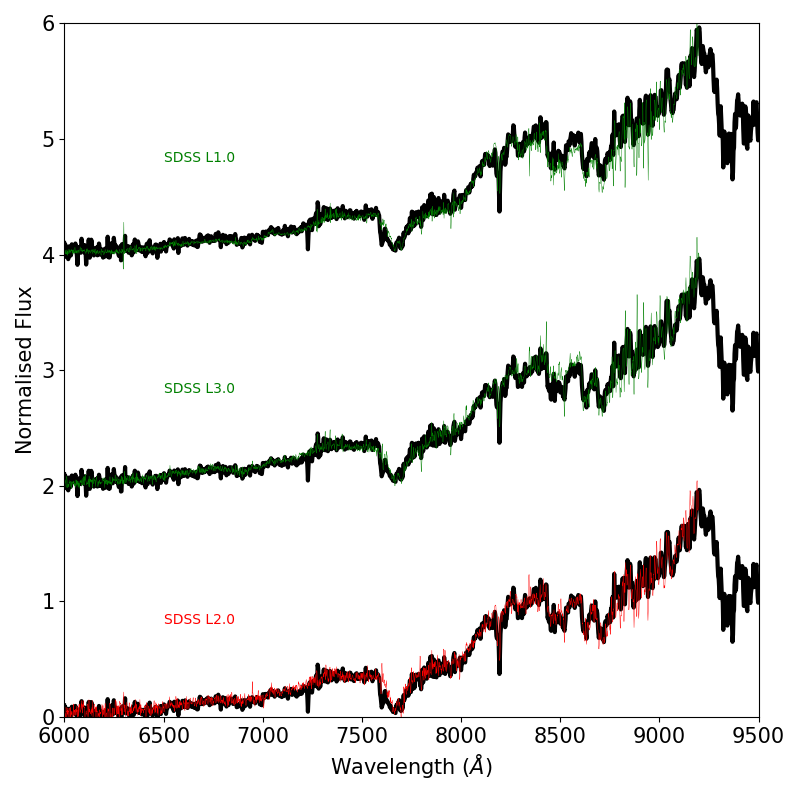}
 \caption{GTC/OSIRIS optical spectra the BD candidate cbd67 (black thik line) compared with L dwarf templates from SDSS (green and red lines).
}
 \label{fig_ComaBer:plot_spec_SDSS}
\end{figure}
\subsection{Spectral typing}
\label{ComaBer:GTCspectro_SpT}
%
 We normalised the OSIRIS spectra between 8240\AA{} and 8260\AA{} to compare with reference ultracool dwarfs of known spectral type because this wavelength range is located in a pseudocontinuum region in late-M dwarfs \citep{martin96}.   

Spectral types were assigned to the targets by visual comparison with 
field late-M and early-L dwarfs from the SDSS database \citep{bochanski07a,schmidt10b}. 
A comparison of the spectra of one of our targets with the SDSS templates is shown in Fig.\ \ref{fig_ComaBer:plot_spec_SDSS}. The spectral types adopted in this work, and those from the literature for 
the sources in common, are listed in Table \ref{tab_ComaBer:spt}.
We infer an optical spectral type of M9.0$\pm$0.5 for cbd34(\,=\,T176) fully consistent
with the infrared spectral types derived independently by \citet{casewell14a} and \citet{tang18a}.
This object is the warmest member candidate in our list of targets with GTC/OSIRIS optical spectra.
We classify cbd10 as a M9.5$\pm$0.5 dwarf, which was lacking a spectrum in \citet{casewell14a}.
We classify optically T159 as a L2.0$\pm$0.5 dwarf, again in agreement with the infrared spectral
type in Table 4 of \citet{tang18a}. Its optical spectrum is similar to cbd67, previously classified
as L1 in the near-infrared by \citet{casewell14a}, as well as to ComaBer\,4, 5, and 7 to which we
assign the same spectral type.
The coolest object in our spectroscopic sub-sample is T191, classified optically as a L2.5$\pm$0.5, although we find it to be 1.5 sub-types earlier than the infrared classification of \citet[][their Table 4]{tang18a}.

%
%
\begin{table*}
 \centering
 \caption[]{Spectral types and radial velocities derived in this work. Spectral types from the literature come from 
 \citet{casewell14a} and \citet{tang18a}\@. Targets are ordered from earlier to later spectral subclass. 
}
 \begin{tabular}{@{\hspace{0mm}}l c c c @{\hspace{0mm}}}
 \hline
 \hline
Name        &    SpT    &   SpT  & RVs  \cr
 \hline
            & Literature & This work & km/s \cr
 \hline
cbd34       & M9 & M9.0$\pm$0.5 & $-$2$\pm$38 \cr
cbd10       &    & M9.5$\pm$0.5 & 3$\pm$46 \cr
cbd67       & L1 & L2.0$\pm$0.5 & $-$18$\pm$35 \cr
ComaBer\,4  &    & L2.0$\pm$0.5 & 16$\pm$24 \cr
ComaBer\,5  &    & L2.0$\pm$0.5 & 2$\pm$15 \cr
ComaBer\,7  &    & L2.0$\pm$0.5 & 62$\pm$16 \cr
T159        & L2   & L2.0$\pm$0.5 & 10$\pm$21 \cr
T191        & L4   & L2.5$\pm$0.5 & $-$1$\pm$14 \cr
 \hline
\label{tab_ComaBer:spt}
\end{tabular}
\end{table*}
\begin{table*}
\centering
\caption{Pseudo EWs of atomic lines in targets observed with GTC/OSIRIS.
}
 \begin{tabular}{@{\hspace{0mm}}l c c c c c c c @{\hspace{0mm}}}
 \hline
 \hline
Name  &  H$_\alpha$ & Li\,{\small{I}}  &  Rb\,{\small{I}}  &  Rb\,{\small{I}}  & Na\,{\small{I}}  & Cs\,{\small{I}} & Cs\,{\small{I}}     \cr
 \hline
      & 6562.8\,\AA{} & 6707.8\,\AA{} & 7800.3\,\AA{} & 7947.6\,\AA{} & 8190\,\AA{} & 8521.1\,\AA{} & 8943.0\,\AA{}   \cr
 \hline
cbd34       & -2.6$\pm$0.8 & $<$1.0 & 1.0$\pm$0.3 & 2.0$\pm$0.3 & 5.5$\pm$0.2 & 1.5$\pm$0.5 & 1.8$\pm$0.5   \cr
cbd10       & $>$-2.3 & $<$3.0 & $<$1.0 & 2.2$\pm$0.5 & 4.9$\pm$0.5 & $<$1.0 & $<$0.5  \cr
cbd67       & $>$-2.1 & $<$1.1 & 2.3$\pm$0.6 & 1.9$\pm$0.7 & 5.8$\pm$0.9 & 3.2$\pm$0.6 & 2.9$\pm$0.4  \cr
ComaBer4    & -5.5$\pm$0.8 & $<$3.3 & 4.0$\pm$1.5 & 3.5$\pm$0.5 & 6.7$\pm$0.5 & 2.5$\pm$0.5 & 2.5$\pm$0.5   \cr
ComaBer5    & -2.0$\pm$1.5 & $<$1.5 & 2.4$\pm$0.2 & 1.5$\pm$0.5 & 5.1$\pm$0.2 & 3.1$\pm$0.3 & 1.7$\pm$0.2  \cr
ComaBer7    & $>$-4.0 & $<$3.0 & 5.0$\pm$0.5 & 3.4$\pm$0.2 & 6.1$\pm$0.3 & 2.9$\pm$0.5 & 1.3$\pm$0.3  \cr
T159        & $>$-4.0    & $<$2.0   & 3.1$\pm$0.3 & 2.3$\pm$0.5 & 5.5$\pm$0.3 & 1.8$\pm$0.5 & 1.9$\pm$0.2  \cr
T191        & $>$-1.5  & $<$1.2 & 4.9$\pm$0.4 & 4.6$\pm$0.9 & 4.7$\pm$0.3 & 2.3$\pm$0.7 & 3.0$\pm$0.4  \cr
DENIS1228   & -1.75$\pm$0.25 & 2.6$\pm$0.6 & 4.7$\pm$1.5 & 6.5$\pm$0.5 & 4.9$\pm$0.3 & 4.6$\pm$0.3 & 4.3$\pm$0.2  \cr
 \hline
\label{tab_ComaBer:pew}
\end{tabular}
\end{table*}
 
\subsection{Radial velocity measurements and cluster membership}
\label{ComaBer:RV}

RV measurements were obtained for all the targets from cross correlation with the spectrum of the template DENIS J1228 using the IRAF task {\it fxcor} over the spectral range 8100$-$10000 $\AA$. Error bars were derived from gaussian fits to the correlation function. 
Heliocentric RV corrections were derived with the IRAF task {\it rvcor}. Instrumental zeropoint correction was applied using a radial velocity of 4 km/s for DENIS J1228 \citep{martin97b}. 
The measured radial velocities obtained for our targets and their uncertainties are provided in Table \ref{tab_ComaBer:spt}.
The RV values of all but one of the Coma Ber targets are consistent within the error bars with the mean RV value of the cluster (-0.52 km/s) provided by $Gaia$ DR2 \citep{gaia18}. The exception is ComaBer7, which 
has a larger RV value of 62$\pm$16 km/s, which deviates from the cluster RV by more than 3 times the uncertainty.

Since cbd67 has been deemed not to be a bona-fide member in Coma Ber because it does not have consistent proper motion \citep{tang18a}, and ComaBer7 has a larger RV value than expected for a member in the cluster, 
we do not include these objects in the subset of confirmed cluster members. Therefore, only 
cbd10/34, ComaBer4/5, and T159/191 are retained as bona fide cluster members confirmed by our RV analysis. As a cautionary note, we remind the reader that ComaBer5 does not have membership confirmation by proper motion.

\subsection{Pseudo equivalent width measurements}

\label{ComaBer:pEW}
\begin{figure}
 \centering
  \includegraphics[width=0.4\linewidth, angle=0]{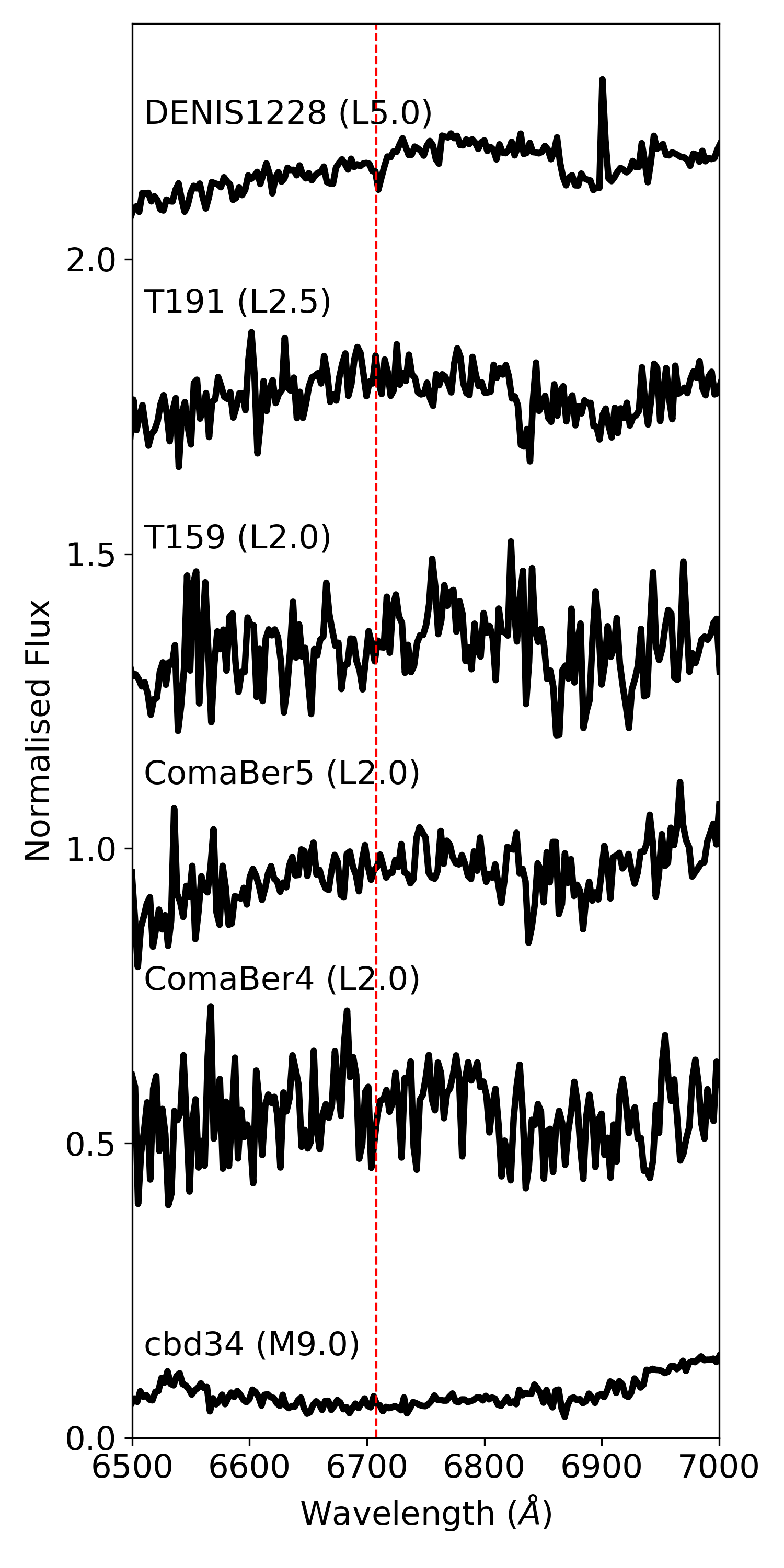}
  \caption{Zoom on the spectral region around the Li resonance doublet at 6707.8\AA{} for Coma Ber objects found by us, and by \citet{casewell06}, \citet{casewell14a}, and \citet{tang18a}, and confirmed as bona-fide cluster members in this work. The location of the Li\,{\small{I}}
resonance doublet is marked with a vertical red line. This feature is clearly detected only in the GTC/OSIRIS spectrum of the lithium L dwarf, DENIS1228\citep{martin97a, martin99a}, but not in the Coma Ber targets. 
}
  \label{fig_ComaBer:plot_spec_Li}
\end{figure}

We searched for the presence of Li\,{\small{I}} in absorption at 6707.8\AA{} in the GTC/OSIRIS spectra. 
A zoom on the Li\,{\small{I}} spectral region is shown in
Fig.\ \ref{fig_ComaBer:plot_spec_Li} for a representative subset of our sample. The Li\,{\small{I}}
resonance doublet clearly stands out from the noise only in our reference lithium L dwarf, DENIS1228 \citep{martin97b, martin99a}, but not in the Coma Ber targets. We did not detect the Li\,{\small{I}} feature in any of the Coma Ber spectra.  Signal to Noise ratios (SNR) in a spectral region adjacent to the Li\,{\small{I}} spectral region were measured with the splot task in IRAF, and are given in Table \ref{tab_ComaBer:log_obs}. This region is the same as that used in \citet{martin18a}. The SNR values measured by us in the Coma Ber targets tend to be slightly lower than to those reported by \citet{martin18a} for Hyades targets, 
and hence our uncertainties in pEW measurements are larger. 

From these data we can only impose upper limits on the Li\,{\small{I}}
pseudo-equivalent width (pEW) of the Coma Ber targets. Those upper limits, together with pEW measurements or upper limits on the emission of H$_\alpha$,
and other atomic lines of interest (Cs\,{\small{I}}, Na\,{\small{I}} and Rb\,{\small{I}}) are provided in Table \ref{tab_ComaBer:pew}. The pEW values were estimated independently by each author using the task splot in the IRAF environment. The mean value of each measurement was adopted and the error bar reflects the dispersion in the values obtained. The pEW values of the Cs\,{\small{I}} and Li\,{\small{I}} features obtained from our spectrum of DENIS1228 are consistent within the measurement uncertainties with those reported by \citet{kirkpatrick99} and \citet{martin99a}. On the other hand, for the Na\,{\small{I}} doublet our value is significantly stronger than the one given by \citet{martin99a}. This could be due to the fact that we partially resolve the Na\,{\small{I}} doublet, whereas it is completely blended in 
the KeckII data presented in \citet{martin99a}, and to contamination by telluric absorption in the red part of the feature.

 Weak H$_\alpha$ in emission is detected in 3 out of the 6 Coma Ber members. The frequency of H$_\alpha$ emitters among Coma Ber M9--L2.5 objects (50\%) is higher than among high-gravity field dwarfs in the same range of spectral subclasses (8 out of 34; 23.5\%) \citep{martin10a} suggesting that chromospheric activity is stronger in the Coma Ber substellar population because it is younger than the field. 

Low-gravity field dwarfs with spectral subclasses from M9 to L0 are thought to have ages younger than about 100 Myr. They have Na\,{\small{I}} pEW values below 4 \AA , and H$_\alpha$ pEW values stronger than pEW = -14 \AA{} \citep{martin10a}. None of the targets observed by us have as weak Na\,{\small{I}} and as strong H$_\alpha$ emission as low-gravity field dwarfs, indicating that they are older, high-gravity ultracool dwarfs. 

%

%
%
%
 
\begin{figure*}
 \centering
  \includegraphics[width=0.7\linewidth, angle=0]{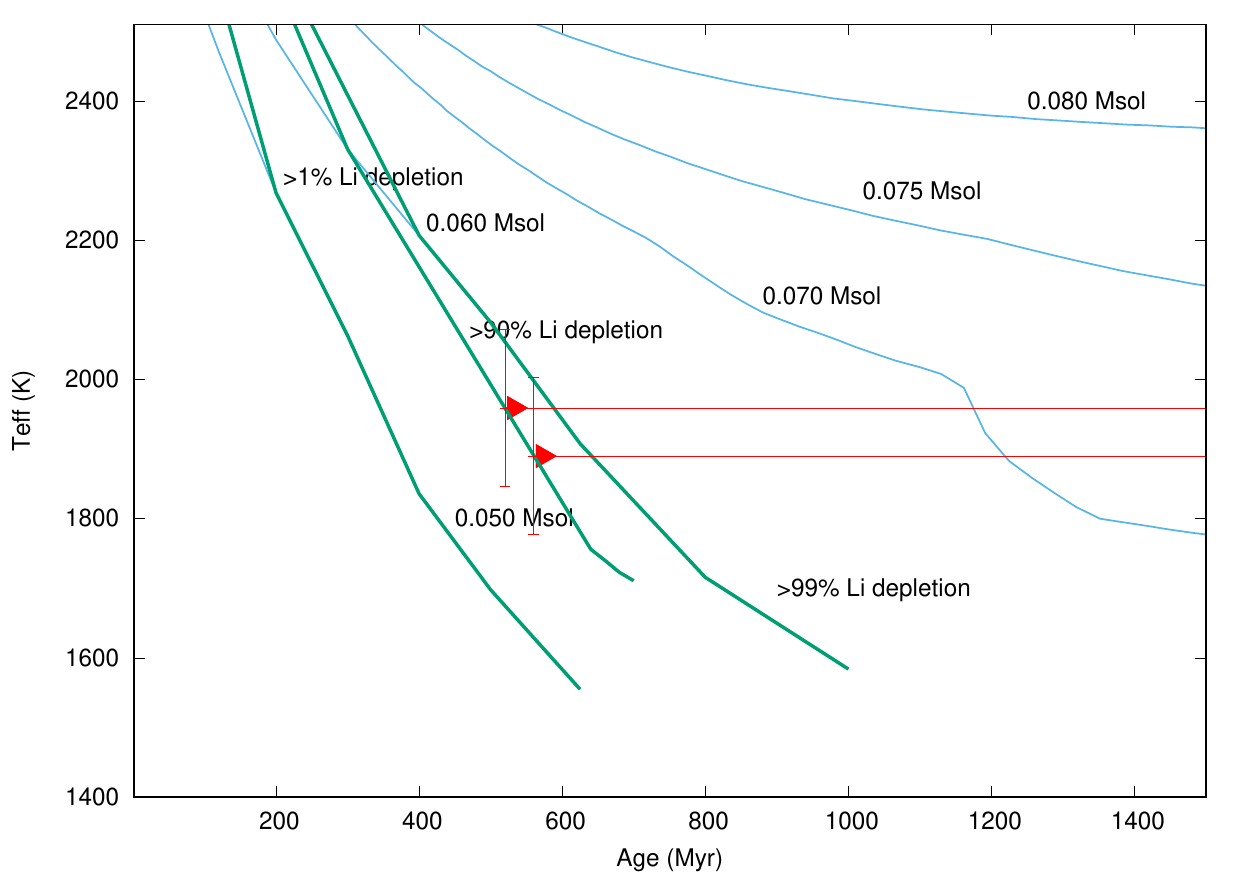}
 \includegraphics[width=0.7\linewidth, angle=0]{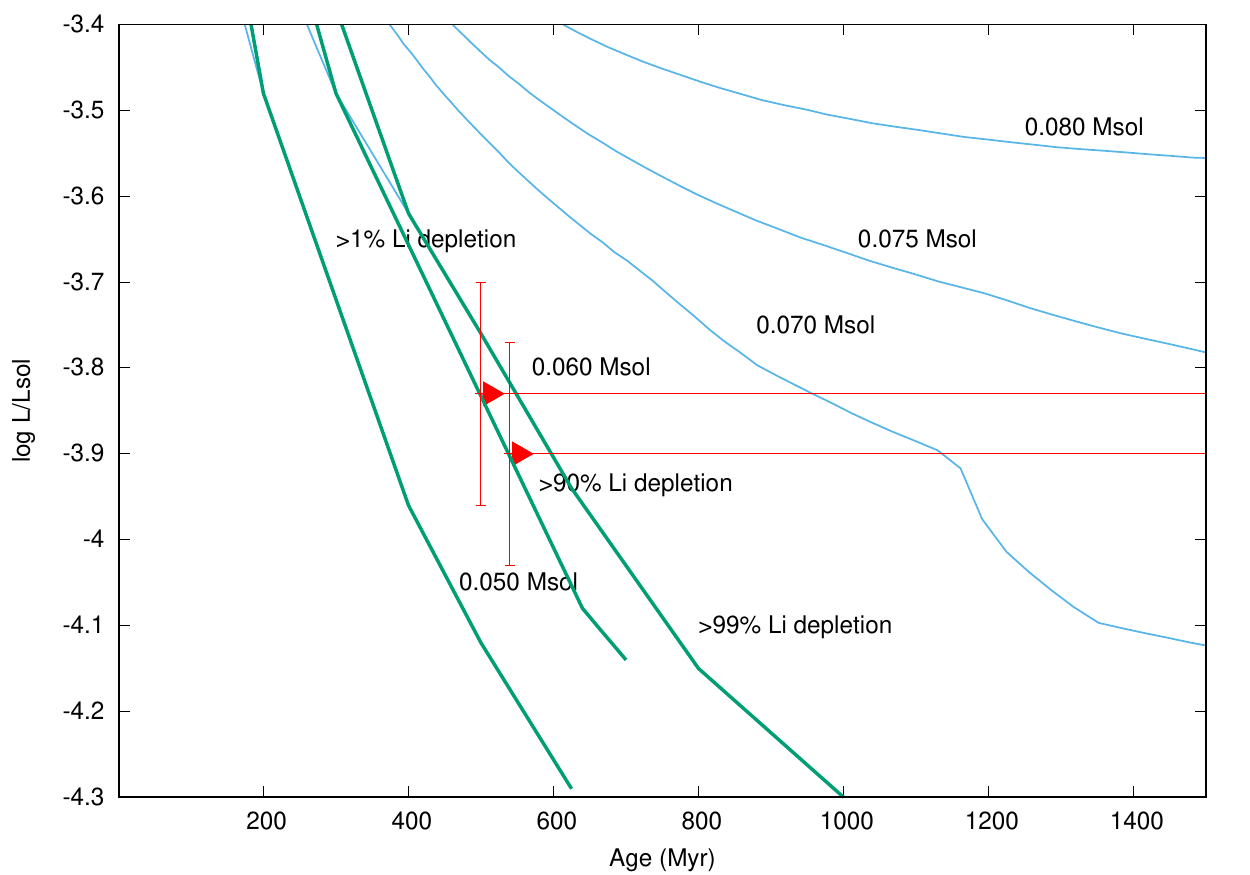}
  \caption{Evolutionary tracks (effective temperature (Teff) and luminosity) from the models by \citet{baraffe15} for VLM stars and BDs as a function of age.
Masses and predicted Li depletion factors are labeled. 
The non detection of Li in the Coma Ber member candidates studied in this work impose lower limits on the cluster age. The limit coming from the L2 candidates is depicted as the upper red triangle and horizontal line, and the limit obtained from the L2.5 object (T191) corresponds to the lower triangle and horizontal line in red colour. 
}
  \label{fig_ComaBer:plot_age_models}
\end{figure*}
 
\begin{table*}
 \centering
 \caption[]{Basic parameters and Li abundances for RV confirmed substellar members in Coma Ber.
}
 \begin{tabular}{@{\hspace{0mm}}l c c c c c  @{\hspace{0mm}}}
 \hline
 \hline
Name        &    SpT    &  $log$(L$_{\rm bol}$/L$_{\odot}$ & T$_{eff}$ & log N(Li) & Age    \cr
 \hline
            &           &             &   (K)   &             & (Myr)    \cr
 \hline
ComaBer\,4  & L2.0$\pm$0.5 &  -3.83$\pm$0.13 & 1959 $\pm$113 &  $<$2.4 &  $>$480      \cr
ComaBer\,5  & L2.0$\pm$0.5 &  -3.83$\pm$0.13 & 1959 $\pm$113 &  $<$1.8 &  $>$500      \cr
T159        & L2.0$\pm$0.5 &  -3.83$\pm$0.13 & 1959 $\pm$113 &  $<$2.0 &  $>$490      \cr
T191        & L2.5$\pm$0.5 &  -3.90$\pm$0.13 & 1890 $\pm$113 &  $<$1.5 &  $>$550      \cr
 \hline
\label{tab_ComaBer:parameters}
\end{tabular}
\end{table*}
 
\begin{table*}
\centering
\caption{Li depletion boundary (LDB) in open clusters older than 20 Myr, ordered by increasing distance.}
\begin{tabular}{@{\hspace{0mm}}l c c c c  c  c @{\hspace{0mm}}}
\hline
\hline
Name        & d  & SpT  & Age (LDB) & Age (Other) & Mass & Ref. \cr
\hline
            &   pc   &  Li   &  Myr  &  Myr  & Msun  &    \cr
\hline
Hyades  & 47 & L4  & 650$\pm$70 &  570 -- 900 & 0.065 & \citet{martin18a} \cr
Coma Ber  & 87 &  $>$L2.5   & $>$550 &  300 -- 1000 & $<$0.07 & This work \cr
Pleiades  & 130 & M6.5  & 112$\pm$5 &  70 -- 160 & 0.075 & \citet{dahm15} \cr
IC 2391  & 146 & M5  & 50$\pm$5 &  30 -- 75 &  0.12 & \citet{barrado04b} \cr
IC 2602  & 152 & M5.5  & 46$\pm$6 &  25 -- 70 & 0.12 & \citet{dobbie10} \cr
Alpha Per  & 190 & M6.5  & 85$\pm$10 &  50--70 &  0.08 & \citet{barrado04b} \cr
Blanco 1  & 207 & M7  & 126$\pm$14 & 150 -- 500 & 0.072 & 
\citet {Juarez14} \cr
IC 4665  & 385 & M4  & 28$\pm$5 & 30 -- 100 &  0.24 & \citet {manzi08} \cr
NGC 2547  & 430 & M4  & 35$\pm$4 & 20 --80 & 0.17 & \citet{jeffries05}\cr
\hline
\label{tab_ComaBer:LDB}
\end{tabular}
\end{table*}

\section{Lithium depletion and the age of Coma Ber}
\label{ComaBer:Age_Li}

The upper limits to the pEW of the Li\,{\small{I}}
resonance doublet reported in Table \ref{tab_ComaBer:pew}, were converted to Li abundances using the method described in \citet{martin18a}, and the same initial Li abundance of log N(Li)$=$3.3 as in the Hyades, which is thought to be representative of the cosmic Li content in newly born stars \citep{martin94a,martin97a}. 

Since Coma Ber is not much younger than the Hyades, it is likely that the L-type members are BDs, but this is not guaranteed for the late-M mmebers. In the following discussion, we retain only the bona-fide substellar cluster members, i.e., the L-type brown dwarfs confirmed by us with radial velocities. As discussed by \citet{martin18a}, it is well justified to use the calibrations for field L dwarfs of known distance to calculate the bolometric luminosities ($log$(L$_{\rm bol}$/L$_{\odot}$) and effective temperatures (T$_{eff}$) from the spectral types obtained by us for the cluster members. The values obtained using the calibrations from \citet{filippazzo15} are provided in Table \ref{tab_ComaBer:parameters}. 

The lack of Li detection implies that the Coma Ber BDs have depleted significant amounts of this light element, which in turn can be used to derive constraints on the age
of the cluster using evolutionary models. In our previous work in the Hyades \citep{martin18a},
we found that there is good agreement between the Li depletion age constraints and other parameters
such as luminosity and temperature for the models by \citep{baraffe15}, and thus we adopt the same models
in this work, as illustrated in Fig.\ \ref{fig_ComaBer:plot_age_models}.
 From these models, and the Li depletion factors obtained among the substellar members in Coma Ber, we infer a lower limit on the cluster age of 550 Myr. The limit obtained from the L2.5 member (T191) is 50 Myr more stringent than from the L2.0 members.  
Age estimates for Coma Ber have ranged from 300 Myr to 1000 Myr \citep{tsvetkov89a}. For example, \citet{kraus07d} used an age of 400 Myr in their study of the low-mass population. 
The lower limit on the age derived in this work restricts the allowed 
range of ages for Coma Ber to lie between 550 Myr and 1000 Myr, and
hence we propose a revised age estimate of 780$\pm$230 Myr, which is consistent with the age estimates in the range 700 -- 800 Myr proposed 
by \citet{tang18a, tang19a} from analysis of a few post-main sequence stars.

A more precise age estimate for this cluster could be derived from Li detection
in members cooler than those studied here. We plan to carry out a search for fainter Coma Ber members using the Euclid wide survey in the framework of the project Independent Legacy Science on ultracool dwarfs, which was selected by ESA in 2012. According to the Euclid specifications, the wide survey should reach about 5 mag. deeper in optical and near-infrared passbands than the surveys that we have used in this work.
If the age of Coma Ber is as old or older than that of the Hyades, the LDB should be located
at spectral type later than L3\@. Li has been detected in L3.5--L5 members in the Hyades \citep{martin18a,lodieu18b}.
Similar observations in L dwarfs in Coma Ber would be challenging but not unfeasible, and they are critical to refine the age of Coma Ber using the substellar Li boundary method.

Coma Ber is the second open cluster closest to us, and also the second one known for which the LDB age is older than 500 Myr. In Table \ref{tab_ComaBer:LDB} we put our LDB results for Coma Ber in context with previous results for other open clusters older than 20 Myr. Note that the results reported here, together with those in the Hyades, have significantly extended the range of applicability of the LDB method. 

As can be seen in Table \ref{tab_ComaBer:LDB}, where the spectral type of the objects where Li reemerges in different open clusters is listed in column 3, the LDB evolves from mid-M spectral types at very young ages (20--50 Myr) to mid-L dwarfs at more mature ages (600--700 Myr). Since BDs keep cooling down with increasing time, the LDB method can still be applied to even older open clusters, moving groups, associations or multiple stellar systems, although it will be harder because the substellar-mass members are much fainter than in younger open clusters, and Li may be locked up in molecules in T dwarfs.
In Coma Ber we are already reaching the practical limit of what can be done with a 10-m class telescope in a reasonable amount of observing time. The new generation of larger optical telescopes will be needed to extend this work to more distant and/or older clusters such as, for example, Praesepe, where BD candidates have already been identified \citep{magazzu98, wang11}, and which has an age similar to Coma Ber although it is located more than twice further away \citep{2019Lodieu}.

The LDB method is orthogonal to other dating methods such as isochrone fitting and gyrochronology, and tends to provide rather precise values that could be useful to constrain the wider range of estimates obtained from different stellar clocks. The validity of the LDB now extends for over a factor of 3 in mass and over a factor of 30 in age, and it provides a promising method to calibrate traditional models of evolved intermediate-mass stars, low-mass pre-main sequence stars, and young main sequence stars.

%
%
\begin{acknowledgements}
Based on observations made with the Gran Telescopio Canarias (GTC), installed in the Spanish Observatorio del Roque de los Muchachos of the Instituto de Astrof\'isica de Canarias, in the island of La Palma (programmes GTC47-18A and GTC55-19A).
ELM, NL, and VJSB were financially supported by the Ministerio de Economia y Competitividad and the Fondo Europeo de Desarrollo Regional (FEDER) under grants AYA2015-69350-C3-1-P and 
AYA2015-69350-C3-2-P\@. 
This research has made use of the Simbad and Vizier databases, operated
at the centre de Donn\'ees Astronomiques de Strasbourg (CDS), and
of NASA's Astrophysics Data System Bibliographic Services (ADS).
This research has also made use of some of the tools developed as part of the
Virtual Observatory.
This work has made use of data from the European Space Agency (ESA) mission
{\it Gaia} (\url{https://www.cosmos.esa.int/gaia}), processed by the {\it Gaia}
Data Processing and Analysis Consortium (DPAC,
\url{https://www.cosmos.esa.int/web/gaia/dpac/consortium}). Funding for the DPAC
has been provided by national institutions, in particular the institutions
participating in the {\it Gaia} Multilateral Agreement.\\
Funding for the Sloan Digital Sky Survey IV has been provided by the Alfred P. Sloan Foundation, the U.S. Department of Energy Office of Science, and the Participating Institutions. SDSS-IV acknowledges
support and resources from the Center for High-Performance Computing at
the University of Utah. The SDSS web site is www.sdss.org.
SDSS-IV is managed by the Astrophysical Research Consortium for the 
Participating Institutions of the SDSS Collaboration including the 
Brazilian Participation Group, the Carnegie Institution for Science, 
Carnegie Mellon University, the Chilean Participation Group, the French Participation Group, 
Harvard-Smithsonian Center for Astrophysics, 
Instituto de Astrof\'isica de Canarias, The Johns Hopkins University, 
Kavli Institute for the Physics and Mathematics of the Universe (IPMU) / 
University of Tokyo, Lawrence Berkeley National Laboratory, 
Leibniz Institut f\"ur Astrophysik Potsdam (AIP),  
Max-Planck-Institut f\"ur Astronomie (MPIA Heidelberg), 
Max-Planck-Institut f\"ur Astrophysik (MPA Garching), 
Max-Planck-Institut f\"ur Extraterrestrische Physik (MPE), 
National Astronomical Observatories of China, New Mexico State University, 
New York University, University of Notre Dame, 
Observat\'ario Nacional / MCTI, The Ohio State University, 
Pennsylvania State University, Shanghai Astronomical Observatory, 
United Kingdom Participation Group,
Universidad Nacional Aut\'onoma de M\'exico, University of Arizona, 
University of Colorado Boulder, University of Oxford, University of Portsmouth, 
University of Utah, University of Virginia, University of Washington, University of Wisconsin, 
Vanderbilt University, and Yale University.
This publication makes use of data products from the Two Micron All Sky Survey, which is a joint 
project of the University of Massachusetts and the Infrared Processing and Analysis 
Center/California Institute of Technology, funded by the National Aeronautics and Space 
Administration and the National Science Foundation.\\
The UKIDSS project is defined in \citet{lawrence07}. UKIDSS uses the UKIRT Wide Field Camera 
\citet[WFCAM;][]{casali07}. The photometric system is described in \citet{hewett06}, and the 
calibration is described in \citet{hodgkin09}. The pipeline processing and science archive are 
described in Irwin et al.\ (2009, in prep) and \citet{hambly08}.\\
This publication makes use of data products from the Wide-field Infrared Survey Explorer, which 
is a joint project of the University of California, Los Angeles, and the Jet Propulsion 
Laboratory/California Institute of Technology, and NEOWISE, which is a project of the Jet 
Propulsion Laboratory/California Institute of Technology. WISE and NEOWISE are funded 
by the National Aeronautics and Space Administration. \\
The Pan-STARRS1 Surveys (PS1) and the PS1 public science archive have been made possible 
through contributions by the Institute for Astronomy, the University of Hawaii, the Pan-STARRS 
Project Office, the Max-Planck Society and its participating institutes, the Max Planck Institute 
for Astronomy, Heidelberg and the Max Planck Institute for Extraterrestrial Physics, Garching, 
The Johns Hopkins University, Durham University, the University of Edinburgh, the Queen's 
University Belfast, the Harvard-Smithsonian Center for Astrophysics, the Las Cumbres Observatory
Global Telescope Network Incorporated, the National Central University of Taiwan, the Space 
Telescope Science Institute, the National Aeronautics and Space Administration under 
Grant No.\ NNX08AR22G issued through the Planetary Science Division of the NASA Science 
Mission Directorate, the National Science Foundation Grant No. AST-1238877, the University 
of Maryland, Eotvos Lorand University (ELTE), the Los Alamos National Laboratory, and the 
Gordon and Betty Moore Foundation. \\
\end{acknowledgements}
%

%
%
\bibliographystyle{aa}
\bibliography{mnemonic,biblio4} 

\end{document}